\documentclass[journal]{IEEEtran}
\makeatletter
\def\ps@headings{%
\def\@oddhead{\mbox{}\scriptsize\rightmark \hfil \thepage}%
\def\@evenhead{\scriptsize\thepage \hfil \leftmark\mbox{}}%
\def\@oddfoot{}%
\def\@evenfoot{}}
\makeatother
\usepackage{tikz,psfrag,epsfig,algorithmic,algorithm}
\usetikzlibrary{shapes,decorations}
\input{tcilatex}

\begin{document}

\title{Memory-Assisted Universal Compression of Network Flows }

\author{
Mohsen Sardari, Ahmad Beirami, Faramarz Fekri\\
School of Electrical and
Computer Engineering, Georgia Institute of Technology, Atlanta, GA 30332\\
\texttt{Email:}\{mohsen.sardari, beirami, fekri\}@ece.gatech.edu
\vspace{-.3in}

\thanks{This material is based upon work supported by the National Science Foundation under Grant No. CNS-1017234.}
}

\newcommand{\mc}{\mathcal}
\newcommand{\mb}{\mathbf}
\newcommand{\BH}{{\sf bit$\times$hop}}

\maketitle
\thispagestyle{empty}
\pagestyle{empty}
\begin{abstract}
Recently, the existence of considerable amount of redundancy in the Internet traffic has stimulated the deployment of several redundancy elimination techniques within the network. These techniques are often based on either packet-level Redundancy Elimination (RE) or Content-Centric Networking (CCN). However, these techniques cannot exploit sub-packet redundancies. Further, other alternative techniques such as the end-to-end universal compression solutions would not perform well either over the Internet traffic, as such techniques require infinite length traffic to effectively remove redundancy. This paper proposes a memory-assisted universal compression technique that holds a significant promise for reducing the amount of traffic in the networks.
The proposed work is based on the observation that if a source is to be compressed and sent over a network, the associated universal code entails a substantial overhead in transmission due to finite length traffic. However, intermediate nodes can learn the source statistics and this can be used to reduce the cost of describing the source statistics, reducing the transmission overhead for such traffics.
We present two algorithms (statistical and dictionary-based) for the memory-assisted universal lossless compression of information sources. These schemes are universal in the sense that they do not require any prior knowledge of the traffic's statistical distribution. We demonstrate the effectiveness of both algorithms and characterize the memorization gain using the real Internet traces. Furthermore, we apply these compression schemes to Internet-like power-law graphs and solve the routing problem for compressed flows. We characterize the network-wide gain of the memorization from the information theoretic point of view. In particular, through our analysis on power-law graphs, we show that non-vanishing network-wide gain of memorization is obtained even when the number of memory units is a tiny fraction of the total number of nodes in the network. Finally, we validate our predictions of the memorization gain by simulation on real traffic traces.
\end{abstract}

\vspace{-.05in}
\vspace{-.2in}

\section{Introduction}

Several studies have shown that there is significant amount of redundant data in the Internet traffic (see~\cite{Spring2000,SIVA-TECH-REPORT} and references therein). Currently, the redundancy elimination techniques are mostly based on end-to-end caching mechanisms~\cite{CLOUD-CONTROL,Spring2000}. It has been argued that the end-to-end approaches do not efficiently leverage the redundancy in the network~\cite{SIVA-TECH-REPORT,ITW11,DCC12_gain}. End-to-end approaches are incapable of suppressing redundancies existing across multiple connections or source-destination pairs. They even lose opportunities for suppressing within one connection because of issues such as short packet length or fixed granularity. To address these issues, a few recent studies have considered the deployment of redundancy elimination techniques within the network~\cite{anand_sigcomm_09,Siva_ACM}, where the intermediate nodes in the network have been assumed to be capable of caching of the previous communication and data processing. These works studied the network flow reduction via ad-hoc solutions such as deduplication of the repeated segments of the traffic without any connections to the Information Theory. Therefore, several fundamental questions such as providing predictions about the achievable gain of memorization (to be defined in the following section) in networks are left open problems.

Information theory has already established the fundamental limit in the compression of infinite length sequences for the class of universal schemes. A compression scheme is referred as universal if it does not require any prior knowledge about the source (sequence) statistics. Hence, it is clear that from the practical point of view, the universal family is more interesting than the non-universal one. However, as shown in~\cite{ISIT11}, there is a significant overhead penalty (i.e., gap from the asymptotic limit) when we attempt to universally compress finite length sequences. Motivated by this, in~\cite{DCC12_gain}, Beirami and Fekri studied the fundamental gain of memory-assisted universal source coding on a simple source-destination link where both the encoder and the decoder have access to a previously obtained sequence of length $m$ from the same source. This length $m$ sequence is what we consider as memory (stored in memory units) which conveys important information about the source statistics. It turns out that this memory could significantly improve the performance of universal compression for every newly generated sequence from the source. Therefore, closing the gap between universal compression performance of finite length and infinite length sequences~\cite{DCC12_gain}.

This paper consisted of two parts. In the first part, we develop practical algorithms that reconfirm the gain of memory in source coding, i.e., the achievable gain in a source-destination link, and quantify the developed theoretical results by using real traffic traces. We compare two algorithms from two different families of universal data compression algorithms. This provides a foundation for both the future development of efficient memory-assisted source coding schemes and the network compression in the second part.

In the second part, we extend our work to find achievable network-wide gain of memory deployment in terms of the reduction of the total traffic in the network in Internet-like power-law network graphs, where only a selected number of nodes are capable of memorization and data processing, i.e., only a number of nodes in the network are memory units. Our model assumes that the memory nodes memorize the previous communications which have passed through them. We further assume that the memorized content is utilized by the same way as in \emph{memory-assisted source coding}, which in turn, results in the \emph{network flow compression}. Finding the achievable gain entails, first finding the optimal location of the memory units in the network (the memory deployment problem), and then finding the optimal routing algorithm for the memory-assisted network compression, given that the memory units' locations are known.

In previous work~\cite{ISIT11, Allerton10,DCC12_gain}, we established that performing memory-assisted universal compression can result in significant reduction of the communication overhead for finite-alphabet Markov sources. In~\cite{ITW11}, we extended the theoretical result on the achievable gain (denoted by $g$) of memory-assisted compression on a single link to a general network and defined a network-wide gain (shown by $\mathcal{G}$) for a general network topology as a function of the number of memory units deployed in the network. Clearly, the network-wide gain is also a function of the achievable gain $g$ of the source which has a fundamental limit depending on the amount of memory $m$. In~\cite{ITW11}, we investigated the scaling behavior of the network-wide gain with the number of nodes in an Erd\H{o}s-R\'enyi random network graph, where we showed that significant improvement is achievable by using a small fraction of the network nodes as memory units. Further, we demonstrated that the savings would have not been achieved if memory was not deployed in the network, i.e., the gains are fundamental to memory deployment. To the best of our knowledge,~\cite{ITW11} is the first work that studies this problem from an information theoretic point of view and provides results on the fundamental benefit associated with memorization that otherwise would have not been achieved.

The present paper attempts to broaden these results. Specifically, the contributions of this paper are as follows:
\begin{itemize}

\item We propose two different practical algorithms for memory-assisted compression in the network based on i) LZ77 universal compression~\cite{LZ77}, and ii) Context Tree Weighting (CTW) universal compression~\cite{Willems1995}. We then derive their fundamental gains and compare performances.

\item We investigate, from information theoretic point of view, the fundamental benefits of memory deployment in power-law graphs, which is applicable to the Internet graph as a special case.

\item We validate our theoretical results via numerical analysis and simulations on real traffic traces.

\end{itemize}

The rest of this paper is organized as follows. In Sec.~\ref{sec:related-work}, we review the related work and describe the fundamental advantages of our approach with respect to the previous work. In Sec.~\ref{sec:memory-assisted}, we briefly review the memory-assisted compression and propose two algorithms for practical implementation. In Sec.~\ref{sec:RPLG-model} we investigate the gain of memory in compression of network flow. In Sec.~\ref{sec:routing}, we investigate the challenges memory units in network impose on shortest path routing algorithms and validate our theoretical results via simulations. Finally, Sec.\ref{sec:conclusion} concludes the paper.


\vspace{-.08in}
\section{Related Work}
\label{sec:related-work}

Recently, there has been a lot of attention regarding the efficient utilization of memory units inside network. Most noticeable are the works on packet-level redundancy elimination~\cite{anand_sigcomm_09} and Content-Centric Networking (CCN)~\cite{Named_Data_Networking,Jacobson2009}. The CCN advocates segmenting data into individually addressable pieces, and propose an architecture where individually addressable data segments can be cached in the network. However, There are several fundamental differences between our work on the memory-assisted compression of network flows and the previous research on networking named content.

The first difference is that our approach deals with the data itself, as opposed to the content name.
As a simple example, two independent servers generating the same content but with different names would still be able to leverage the memory-assisted compression, but not the CCN. 
The second difference is that CCN has a fixed granularity of a packet, whereas one of the core features of compression algorithms is their flexibility to find redundancy in the data stream with arbitrary granularity. In fact, it is suggested that packet level caching, which most of the current techniques are approximately reduced to, offers negligible benefits for typical Internet traffic~\cite{SIVA-TECH-REPORT}, due to this predefined fixed granularity.

Another line of work considers the benefits of the deployment of packet-level redundancy elimination in the network~\cite{anand_sigcomm_09,Anand2008}. The redundancy elimination identifies the largest chunk of data that appears in memory and replaces it with a pointer. This algorithm suffers from two issues: 1) The implementation complexity is not linear with the input size, and hence is not scalable for larger packet sizes. 2) The redundancy elimination works well only when the redundancy across the packets is so high that a large chunk of the packet is simply the repetition of a previously communicated packet. However, as we explain in Sec.~\ref{sec:memory-assisted}, there is considerable amount of redundancy in the contents, although the redundant data are not mere repetitions of the previous data chunk, which cannot be exploited using the packet-level redundancy elimination method. 3) If the redundant chunk is repeated with a low frequency such that it does not repeat in the memory window, it may not be detected and may not be removed.
Further, our approach treats the problem of the redundant traffic at the fundamental level and attempts to have an information theoretic view. It formulates the problem as a memory-assisted compression problem. Memory-assisted compression will set new achievable rate regions and scaling trends, e.g., versus number of memory units, number of clients, etc. At the protocol level, memory-assisted compression requires new types of compression techniques that would take into account what is already memorized in the memory units. This type of problem is unique from what is already investigated in the information theory in the context of the source compression. 

While relevant, the memory-assisted source coding problem is different from those addressed by distributed source compression techniques (i.e., the Slepian Wolf problem) that target multiple correlated sources sending information to the same destination~\cite{slepian_wolf,Mina_TCOM}. In the Slepian-Wolf, the gains are achievable in the asymptotic regime. Further, the memorization of a sequence that is statistically independent of the sequence to be compressed can result in a gain in memory-assisted compression whereas, in the Slepian-Wolf problem, the gain is due to the bit by bit correlation between the two sequences.

\vspace{-.05in}
\section{Algorithms for Memory-Assisted Compression}
\vspace{-.05in}
\label{sec:memory-assisted}

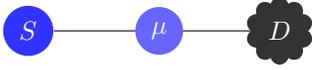
\begin{figure}
\vspace{.2in}
\centering
\begin{tikzpicture}
\draw (0,0) node[anchor=east, circle, fill=blue!80, text=white]{$S$}
-- (1.4,0) 	node[circle, fill=blue!60, text=white]	(m){ $\mu$}
-- (3,0) 	node[cloud, fill=black!80, text=white]	(D){{$D$}};
\end{tikzpicture}
\caption{The basic source, memory, destination configuration. The node $D$ represents a set of clients receiving data from $S$.}
  \label{fig:S-D}
 \vspace{-.2in}
\end{figure}

{\em Memory-Assisted Source Coding:} Consider a source node $S$ which generates content to be delivered to a destination (client) node $D$ (which may be a subnetwork) connected to $S$ through memory node $\mu$, as shown in Fig.~\ref{fig:S-D}. 
Let $X^n$ denote a sequence of length $n$ as the source output.
Let $\mb{E} l_n(X^n)$ denote the expected length resulting from the universal compression of $X^n$. Further, the client nodes in $D$ request various sequences from the source over time.

Consider two coding scenarios: 1) Universal compression without using memory (Ucomp), 2) Memory-assisted source coding (UcompM). In Ucomp, the intermediate nodes simply forward source packets to the subnetwork $D$. As such, compression takes place in the source and decompression is performed in the destination. Assuming a universal compression at source, $\mb{E} l_n(X^n)$ would be the length of the compressed sequence, which has to travel within the network from $S$ to the destination $D$ through the intermediate node $\mu$. Since every client in $D$ requests a different sequence $X^n$ over time, the source must encode each sequence $X^n$ independently and route through $\mu$. Now, consider the second scenario in which the intermediate node $\mu$, while serving as an intermediate node for different contents destined for different clients, also constructs a model for the source $S$. As both source and intermediate nodes are aware of the previous content $X^n$ sent to another client in $D$, they can leverage this knowledge for the better compression of the traffic sent over $S-\mu$ portion of the path. 

Specifically, assume that previous sequences $X^{m_1}, \ldots, X^{m_L}$ are sent from $S$ to clients $D_1, \ldots D_L$ in the subnetwork $D$ via $\mu$. Under UcompM, the node $\mu$ constructs a model for the source $S$ by observing the entire length $m = m_1 + \ldots +m_L$ sequence. Note that forming the source model by node $\mu$ is not a passive storage of the sequences $X^{m_1},\ldots,X^{m_L}$. This source model would be extracted differently for different universal compression schemes that we will use as the underlying memory-assisted compression algorithm. Using~\cite{DCC12_gain}, we have proved in~\cite{ISIT11} that UcompM which utilizes the memorized sequences of total length $m$, strictly outperforms Ucomp. This benefit, offered by memorization at node $\mu$, would provide savings on the amount of data transferred on the link $S-\mu$ without incurring any penalty except for some linear computation cost at node $\mu$. Please note that the memorization is used in both the encoder (the source) and the decoder (node $\mu$). Thus, source model is available at both $S$ and $\mu$. 

Let $\mb{E}l_{n| m}(X^n)$ be the expected code length for a sequence of length $n$ and a memorized sequence of length $m$. The fundamental gain of compression $g\left(n,m\right)$ is defined as 
$$
g\left(n,m\right) \triangleq \frac{\mb{E}l_{n}(X^n)}{\mb{E}l_{n|m}(X^n)}.
$$
In other words, $g(n,m)$ is the compression gain achieved by UcompM for the universal compression of the sequence $X^n$ over and above the compression performance that is achieved using the universal compression without memory. The gain $g(n,m)$ is characterized for parametric sources in~\cite{ISIT11}.

From now on, we will refer to those intermediate nodes which construct a source model as memory units. Further, by memory size we mean the total length $m$ of the observed sequences from the source at the memory unit.
To investigate the gain of memorization in the compression of the network flow, we must consider two phases. The first is the memorization phase in which we assume memory units have observed one or multiple sequences of total length $m$ from  the source. This phase is realized in actual communication networks by observing the fact that a sufficient number of clients may have previously retrieved different small to moderate length sequences from the server such that, via their routing, each of the memory units has been able to memorize the source and form a model for it. In the second phase, each client may request (a small to moderate length) content from the source. The memory-assisted source coding is performed in the second phase. 

In the following, we first describe practical algorithms for memory-assisted source coding, by modifying two well-known compression algorithms, namely the statistical compression methods and the dictionary-based compression method. Furthermore, we extend our source model to real Internet traces and characterize the achievable gains $g(n,m)$. In particular, we consider data that was gathered from CNN web server in seven consecutive days. We arbitrarily chose the web server and similar patterns could be found using data from different web servers. Note that these results are provided as a proof of concept for memory-assisted source coding on real Internet traces. From Sec.~\ref{sec:RPLG-model} to the rest, we assume that the fundamental gain of memory-assisted source coding is known and is denoted by $g = g(n, m)$, using which we will derive the achievable network-wide gain.

\vspace{-.08in}
\subsection{Statistical Compression Method: Context Tree Weighting}
The essence of statistical compression methods is to find an estimate for the statistics of the source based on the currently observed sequence or an external auxiliary sequence.
One of the most well-known statistical compression algorithms used in practice is Context Tree Weighting (CTW)~\cite{Willems1995}. 
The context of $X^n$ is defined as the $m$ bits that precede it.
In short, the encoding of every new bit entails: 1) estimating the probability of the bit based on the context, 2) sending the estimated probability to the arithmetic encoder, which encodes the symbol, and 3) updating the context tree with the new data. As expected, the decoding process is similar to the encoding.

The generalization of the CTW encoding/decoding algorithm for the case of memory-assisted compression is immediate. As previously discussed, in memory-assisted compression we assume that a sequence with the true statistics of the source is available both to the decoder (at $\mu$) and the encoder (at $S$). Therefore, by aggregating the data generated at the source and passing through $\mu$, a context tree can be constructed that will be further updated in the compression process.
Note that the source and memory node should always keep the context tree synchronized with each other. In practical settings, a simple acknowledgment mechanism suffices for the context synchronization.  

\begin{figure*}
\begin{center}
  \subfigure[Memory-assisted CTW]{
  \vspace{-.1in}
  \includegraphics[width=.83\columnwidth]{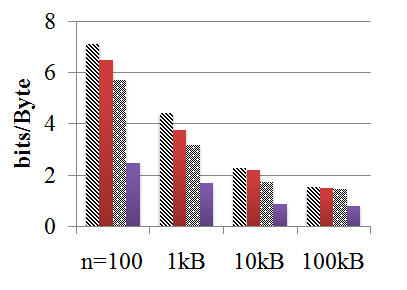}
  \vspace{-.1in}
  \label{fig:ctw}
  }
  \subfigure[Memory-assisted LZ]{
  \vspace{-.1in}
  \includegraphics[width=.83\columnwidth]{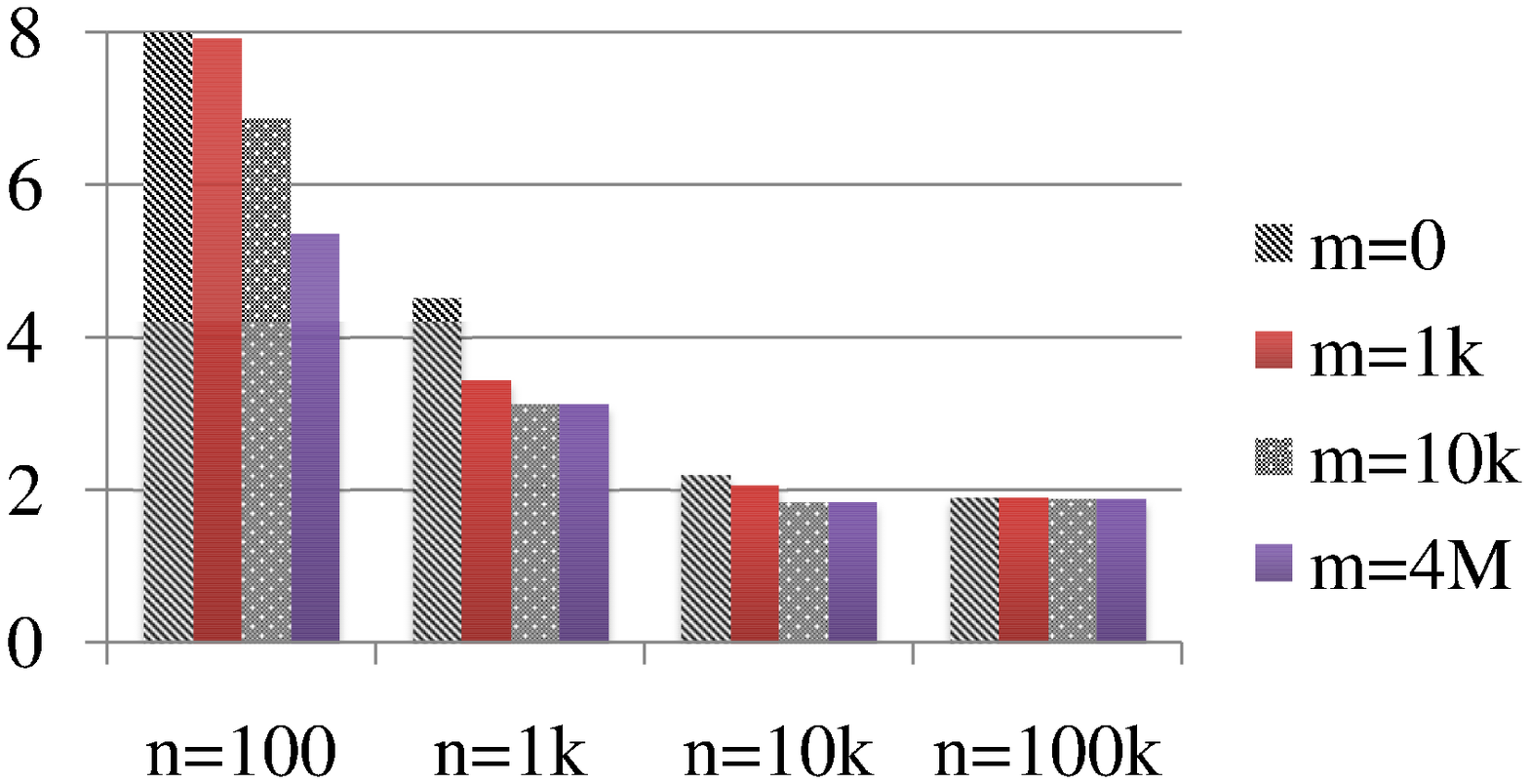}
  \vspace{-.1in}
  \label{fig:lz}
  }
   \subfigure{
  \includegraphics[width=.23\columnwidth]{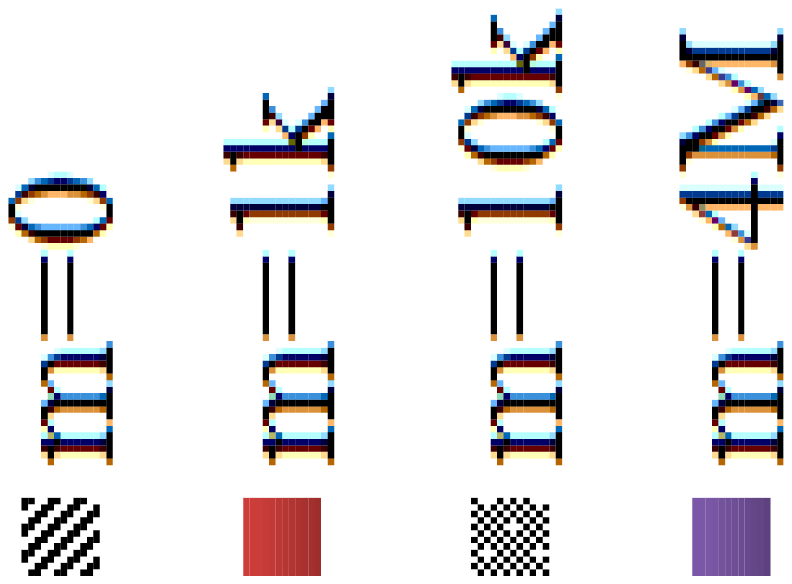}
  }
\end{center}
\vspace{-.1in}
\caption{The compression ratio (bits/Byte) achieved by memory-assisted algorithms.}
\label{fig:memory-assisted gains}
\vspace{-.2in}
\end{figure*}

\vspace{-.08in}
\subsection{Dictionary-based Compression Method: LZ77}
Unlike the statistical compression methods that rely on the estimation of the source statistical parameters, dictionary-based compression methods select sequences of symbols and encode each sequence using a dictionary of sequences that is generally constructed using the previously compressed symbols. 
One of the most efficient dictionary-based methods, that we investigate in this paper, is the LZ77 algorithm~\cite{LZ77}. The principle of LZ77 is to use part of the previously-seen input stream as the dictionary. The encoder maintains a window to the input stream and shifts the input in that window from right to left as strings of symbols are being encoded. 

Our implementation of the memory-assisted LZ77 is based on the open-source DEFLATE algorithm. A sequence of length $m$ is assumed to be available at both encoder and decoder. The previously seen sequence is then used as the common dictionary. The new data to be compressed, is appended to the end of the dictionary at the source and fed to the LZ77 encoder. The output is sent to the decoder. Similarly, the decoder can reconstruct the intended stream by appending the transmitted symbols to the end of the dictionary and perform the LZ77 decoding algorithm. 

\vspace{-.08in}
\subsection{Simulation Results}

We have implemented the memory-assisted CTW and quantized the achievable gains using real Internet traffic data, as shown in Fig.~\ref{fig:ctw}. As expected, the size of the compressed sequence decreases as memory size $m$ increases. For example, for a data sequence of length $n=100$ Bytes, without memory, the compressed sequence has a length of $\approx 87$ Bytes, while using a memory of size $m=4$MB, this data sequences can be compressed to 31 Bytes; almost 3 times smaller. Our simulation results for memory-assisted LZ77 with a window size of 32kB (all seizes are reported in Bytes) and various dictionary sizes are shown in Fig~\ref{fig:lz}.   

\begin{figure}
\begin{center}
  \includegraphics[width=.85\columnwidth]{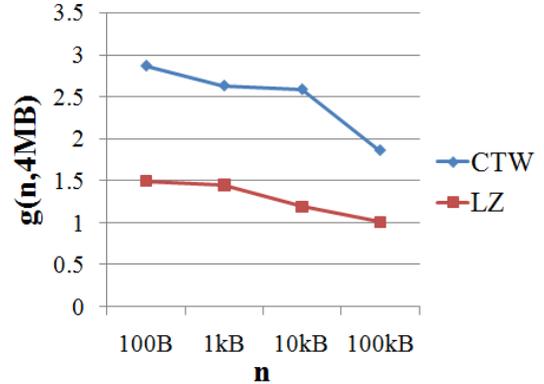}
\end{center}
  \vspace{-.1in}
  \caption{The gain $g$ of memory-assisted compression over Ucomp, for memory size of 4MB for CTW and LZ compression algorithms. This gain is achieved by utilizing memory on top of the gain of conventional compression.}
  \label{fig:gain}
\vspace{-.2in}
\end{figure}

The actual gain of memory-assisted compression $g$ for memory size 4MB is depicted in Fig.~\ref{fig:gain}. Our results suggest that the memory-assisted statistical compression method outperforms the dictionary-based method in both the absolute size of the compressed output and also the gain of memory, i.e., the gain achieved on top of the gain of conventional compression, by utilizing memory.
Both dictionary-based and statistical compressors can achieve the entropy limit for very large input sequences.
However, they perform poorly for short to moderate length sequences.
Therefore, the main advantage of our memory-assisted compression is that it will overcame this limitation by exploiting the available memory.

\vspace{-.05in}
\section{Fundamental Gain of Memory for Flow Compression in Power-law Network Graphs}
\vspace{-.04in}
\label{sec:RPLG-model}
We now extend the basic memory-assisted source coding scenario (Fig.~\ref{fig:S-D}) into network flow compression with memory. The question we try to answer is ``Given the memory-assisted source coding gain $g$, and a number of nodes capable of performing compression, what is the achievable network-wide compression gain of memorization?'' We emphasize again that this network-wide gain $\mc{G}$ is on top of what is achieved if we merely used a compression at the source (without any assistance from memory units).

The answer to this question entails first finding the optimal strategy for deploying the memory units and then investigating the effect of these memory units on the routing algorithms. The latter is important for the numerical evaluation of the network-wide gain as well. In the following, we first introduce the notations involved and describe the memory deployment problem. Then, we discuss the routing problem.

\vspace{-.08in}
\subsection{Notation and Setting}
A network is represented by an undirected graph $G(V,E)$ where $V$ is the set of $N$ nodes (vertices) and $E=\{uv:u,v \in V\}$ is the set of edges connecting nodes $u$ and $v$. We consider a set of memory units $\mathbf{\mu} = \{ \mu_i\}_{i=1}^M$ chosen out of $N$ nodes where every memory node $\mu_i$ is capable of memorizing the communication passing through it. The total size of memorized sequences for each $\mu_i$ is assumed to be equal to $m$.
In this work, we focus on the expected performance of the network by averaging the gain over all scenarios where the source is chosen to be any of the nodes in the network equally at random.
In other words, we assume that the source is located in any node of the network with probability $\frac{1}{N}$. As discussed before, end-to-end compression techniques will only be able to compress the content to a value which may be significantly larger than the entropy of the flow (data sequence).

After the memorization phase, we can assume the constructed source model is available to all memory units. In the second phase, which is the subject of this section, we assume each node in the entire network may request content from the source.  
The above view simplifies our study as we are not concerned with the transition phase during which the memorization is taking place in the memory units. Hence, we can assume that each memory unit will provide the same memory-assisted compression gain of $g$ on the link from the origin node of the flow to itself.

Consider the outgoing traffic of one of the nodes in the network, named as $S$, with the set of its destinations $\mathbf{D}= \{D_i\}_{i=1}^{N-1}$ each receiving different instances of the source sequence originated at $S$. Let $f_D$ be the unit flow from $S$ destined to $D \in \mathbf{D}$. The distance between any two nodes $u$ and $v$ is shown by $d(u,v)$. The distance is measured as the number of hops in the lowest cost path between the two nodes. As we will see later, introducing memories to the network will change the lowest cost paths, as there is a gain associated with the $S-\mu$ portion of the path. Therefore, we have to modify paths accounting for the gain of memories. Accordingly, for each destination $D$, we define \emph{effective walk}, denoted by $W_D = \{S,u_1,\ldots,D\}$, which is the ordered set of nodes in the modified (lowest cost) walk between $S$ and $D$. Finding the shortest walk is the goal of routing problem with memories (Sec.~\ref{sec:routing}). 

We partition the set of destinations as $\mathbf{D} = \mathbf{D}_1 \cup \mathbf{D}_2$, where $\mathbf{D}_1=\{D_i:\exists {\mu}_{\scriptscriptstyle D_i} \in W_{D_i}\}$ is the set of destinations observing a memory in their effective walk, and $\mu_{\scriptscriptstyle D_i} = \arg \min_{\mu \in \mathbf{\mu}} \{\frac{d(S,{\mu})}{g} + d({\mu}, {\scriptstyle D_i})\}$. The total flow $\mathcal{F}_S$ of node $S$ is then defined as
\begin{equation}
\label{eq:totalflow}
\begin{array}{lcl}
\mathcal{F}_S &\triangleq&
\sum_{D_i \in \mathbf{D}_1}\left(\frac{f_{D_i}}{g}d(S,{\mu}_{\scriptscriptstyle {D_i}}) + f_{\scriptscriptstyle D_i}d({\mu}_{\scriptscriptstyle D_i},{\scriptstyle D_i} ) \right) +\\
&&\sum_{D_j \in \mathbf{D}_2} f_{D_j} d(S,D_j)
\end{array}						
.\end{equation}
Using~\eqref{eq:totalflow}, we define $\hat{d}_D$, called the \emph{effective distance} from $S$ to $D$, as
\begin{equation}
\label{eq:effective_dist}
\hat{d}_D = \left\{
\begin{array}{ll}
\frac{d(S,{\mu}_{\scriptscriptstyle D})}{g} + d({\mu}_D, {\scriptstyle D})	& D\in  \mathbf{D}_1\\
d(S,D) 							& D\in  \mathbf{D}_2
\end{array}
\right.
.\end{equation}
In short, the effective distance is the distance when memory-assisted compression is performed and hence the gain $g$ applies. By definition, $\hat{d}_D \leq d(S,D)~ \forall D$.

In a general network, we define a generalized network-wide gain of memory deployment as a function of memorization gain $g$, as follows:
\begin{equation}
\label{eq:network-wide-gain}
\mathcal{G}(g)\triangleq\frac{\sum_{S\in V}\mathcal{F}^0_S}{\sum_{S\in V}\mathcal{F}_S}=\frac{\sum_{S\in V}\sum_{D\in \mathbf{D}} d(S,D)}{\sum_{S\in V}\sum_{D\in \mathbf{D}} \hat{d}_D}
,\end{equation}
where $\mathcal{F}^0_S$ is the total flow in the network by node $S$ without using memory units, i.e., $\mathcal{F}^0_S = \sum_{D\in \mathbf{D}} d(S,D)$. In other words, $\mathcal{G}$ is the gain achieved by memory-assisted scheme on top of what could be saved by an end-to-end compression scheme at the source without using memory. Alternatively, $\mc{G}(g)$ can be rewritten as
$$
\frac{ \sum_{S\in V}\sum_{D\in \mathbf{D}} d(S,D)\mb{E}l_n(X^n)}{ \sum_{S\in V}\sum_{D\in \mathbf{D}} \left[d(S,{\mu}_{\scriptscriptstyle D})\mb{E}l_{n|m}(X^n) + d({\mu}_D, {\scriptstyle D})\mb{E}l_n(X^n)\right]}.
$$

\vspace{-.05in}
\subsection{Memory Deployment in Random Power-law Graphs}
Given a set of memories $\mathbf{\mu}$, the goal of the memory deployment is to maximize $\mathcal{G}$, by optimally deploying those memories. In~\cite{Krishnan2000}, authors studied a related problem in the context of caching to maximize the cache hit-rate in the network. There, they showed that the cache placement problem on a general graph is an NP-hard problem. 
It is straightforward to draw analogies between our memory deployment problem and the cache placement problem, which leads to the fact that the memory deployment problem in a general graph is NP-hard. But, if we limit the deployment problem to certain families of graphs, we can find theoretical solutions that provide insight and enable us to predict the achievable gains.
  
In order to extend our study to the analysis of the achievable gain $\mathcal{G}(g)$ in general networks, we consider the memory deployment gain in the network graphs that follow the power-law degree distribution~\cite{chung-diameter}. The power-law graphs are particularly of interest because they are one of the useful mathematical abstraction of real world networks, such as the Internet and social networks. In power-law graphs, the number of vertices whose degree is $x$, is proportional to $x^{-\beta}$, for some constant $\beta > 1$. For example, the Internet graphs have powers ranging from 2.1 to 2.45~\cite{Albert1999,Faloutsos1999,Broder2000}. 
Accordingly, in the rest of this paper we specifically direct our attention to power-law graphs with $ 2<\beta<3 $ (which include the models for the Internet graph), and provide results for memory deployment on such network graphs. 
We wish to study the behavior of the network-wide gain in Internet-like power-law graphs, as a function of the number of the memory units and their locations.


{\em Random Power-law Graph Model:} A power-law graph is an undirected, unweighted graph whose degree distribution approximates a power law with parameter $\beta$. Basically, $\beta$ is the growth rate of the degrees. In order to generate a random graph that has a power-law degree distribution, we consider the Fan-Lu model~\cite{chung-book}. In this model, we first associate with each vertex, the expected degree of that vertex. The Random Power-Law Graph (RPLG), with parameter $\beta$, is defined as follows:
\begin{definition}[Definition of $G(\beta)$]
Consider the sequence of the expected degrees $\mathbf{w} = \{w_1,w_2,\ldots,w_N\}$, and let $\rho = 1/\sum w_i$. For every two vertices $v_i$ and $v_j$, the edge $v_iv_j$ exists with probability $p_{ij}=w_iw_j\rho$, independent of other edges. If
\begin{equation}
\label{eq:weight}
w_i = ci^{-\frac{1}{\beta-1}} \text{ for } i_0\leq i \leq N+i_0,
\end{equation}
then graph $G$ constructed with such an expected degree sequence is called a RPLG with parameter $\beta$. Here, the constant $c$ depends on the average expected degree $\bar{w}$, and $i_0$ depends on the maximum expected degree $\Delta$. That is, 
\begin{equation*}
\left\{
\begin{array}{lcl}
c	&=&	\frac{\beta-2}{\beta-1}\bar{w}N^{\frac{1}{\beta-1}},\\
i_0	&=&	N\left(\frac{\bar{w}(\beta-2)}{\Delta(\beta-1)}\right)^{\beta-1}.
\end{array}
\right.						
\end{equation*}
\label{def:graph}
\end{definition} 

With the definition above, it is not hard to show that the expected number of vertices of degree $x$ in $G(\beta)$ is $\approx x^{-\beta}$. In~\cite{chung-book}, authors showed that for a sufficiently large RPLG, if the expected average degree of $G(\beta)$ is greater than 1, then $G(\beta)$ has a unique giant component (whose size is linear in $N$), and all components other than the giant component have size at most $O(\log N)$, with high probability. Since we only consider connected networks, we will focus on the giant component of $G(\beta)$ and ignore all sublinear components. Further, by a slight abuse of the notation, by $G(\beta)$ we refer to its giant component. 
Next, we briefly describe as to how the structure of RPLG provides insight about the efficient placement of memory units.  

Although memory deployment problem in a general graph is a hard one, the RPLG with parameter $2<\beta<3$ has a certain structure that leads us to finding a very good deployment strategy. The RPLG can be roughly described as a graph with a dense subgraph, referred to as the \emph{core}, while the rest of the graph (called periphery) is composed of tree-like structures attached to the core. Our approach to solve the memory deployment problem is to utilize this property and size the core of $G(\beta)$ and show that almost all the traffic in $G(\beta)$ passes through the core. We propose to equip all the nodes in the core with memory and hence almost all the traffic in $G(\beta)$ would benefit from the memories. The number of memories should be such that the network-wide gain is greater than 1 as $N\rightarrow \infty$. Showing that almost all the traffic goes through the core guarantees that $\mathcal{G}>1$ as shown in Lemma~\ref{lem:memory-on-path} below. This way, we find an upper bound on the number of memories that should be deployed in a RPLG in order to observe a network-wide gain of memory. We will also verify that the number of memory units does not have to scale linearly with the size of the network to achieve this gain.

From Def.~\ref{def:graph}, we note that the nodes with higher expected degrees are more likely to connect to each other and also other nodes. Therefore, we expect more traffic to pass through these nodes. In our case, we are looking to size the core, i.e., find the number of high degree nodes such that almost all the traffic in the graph passes through them. Theorem~\ref{thm:core} below is our main result regarding the size of the core:
\begin{theorem}
\label{thm:core}
Let $G(\beta)$ be an RPLG. In order to achieve a non-vanishing network-wide gain $\mathcal{G}$, it is sufficient to deploy memories at nodes with expected degrees greater than $lw_{\min}$, where $l$ is obtained from
\begin{equation}
\label{eq:l-beta}
l^{3-\beta}-\frac{1}{\bar{w} \gamma}=0
,\end{equation}
and the constant $\gamma$ is equal to $(1- \frac{1}{\beta-1})^2 \frac{\beta-1}{3-\beta}$. The set of nodes with expected degree greater than $lw_{\min}$ is defined as core: $\mathcal{C}=\{u|w_u > lw_{\min}\}$.
\end{theorem}
Proof of the Theorem~\ref{thm:core} follows from the lemmas below.
\begin{lemma}
\label{lem:memory-on-path}
Let $d$ be the distance between the nodes $A$ and $B$. Let $\mu$ denote a memory unit fixed on the shortest path between $A$ and $B$, with distance $d'$ from $A$, i.e., the distance between $\mu$ and $B$ is $d-d'$. If the fundamental gain of memory-assisted compression is $g>1$, then $\mathcal{G}>1$.
\end{lemma}
\begin{IEEEproof}
If there was no memory on the path, we had one unit of flow from $A$ to $B$ and one unit of flow for $B$ to $A$. Therefore, $\mathcal{F}^0 = 2d$. When memory-assisted compression is performed, the flow going from $A$ to $B$ is reduced to $\frac{d'}{g}+(d-d')$. Similarly, the flow going from $B$ to $A$ is $\frac {d-d'}{g}+d'$. Therefore, $\mathcal{F} = \frac{d'}{g}+(d-d') + \frac {d-d'}{g}+d'$ and thus
\[
\mathcal {G}(g) = \frac{2d}{\frac{d'}{g}+(d-d') + \frac {d-d'}{g}+d'}=\frac{2g}{g+1}.
\]
Now, considering that $g>1$, the claim follows.
\end{IEEEproof}

Our approach to find the core is to remove the highest degree nodes from the graph one at a time until the remaining induced subgraph does not form a giant component. In other words, as a result of removing the highest degree nodes, the graph decomposes to a set of \emph{disjoint} islands and hence, we conclude that the communication between those islands must have passed through the core.
Therefore, from Lemma~\ref{lem:memory-on-path}, we conclude that in RPLG, we will have a non-vanishing network-wide gain if we choose the core sufficiently big such that the induced periphery of $G(\beta)$ does not have a giant component. The following lemma provides a sufficient condition for not having a giant component in RPLG.
\begin{lemma}[\cite{chung-book}]
\label{lem:giant-component}
A random graph $G(\beta)$ with the expected degrees $\mathbf{w}$, almost surely has no giant components if 
\begin{equation}
\label{eq:nogiantcomponent}
\frac{\sum_i w_i^2}{\sum_i w_i} < 1.
\end{equation}
\end{lemma}

\begin{lemma}
\label{lem:subgraph}
Consider a random graph $G$ with the sequence of the expected degrees $\mathbf{w}$. If $U$ is a subset of vertices of $G$, the induced subgraph of $U$ is a random graph with the sequence of the expected degrees $\mathbf{w'}$, where 
\[
w_i' = w_i\frac{\sum_{v\in U}w_v}{\sum_{v\in G}w_v}
.\]
\end{lemma}
\begin{IEEEproof}
The probability that an edge exists between two vertices of $U$ is equal to the edge connection probability in $G$. Consider a vertex $u$ in $U$. The expected degree of $u$ is
$$
\rho\sum_{v\in U}w_uw_v = w_u \frac{\sum_{v\in U}w_v}{\sum_{v\in G}w_v}.
$$
\end{IEEEproof}

\begin{figure}
\begin{center}
  \includegraphics[width=.85\columnwidth]{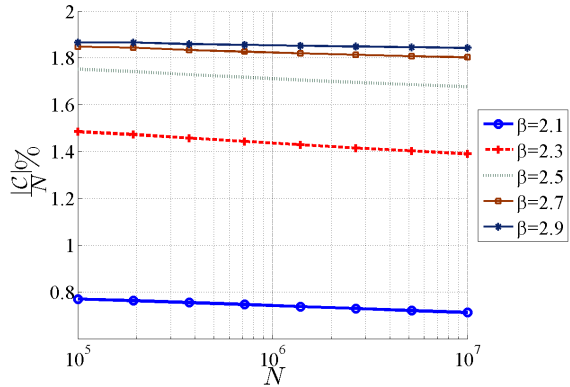}
\end{center}
\vspace{-.1in}
\caption{The scaling of the core size $\frac{|\mathcal{C}|}{N}\times 100$ versus $N$ for different $\beta$'s.}
\label{fig:U_l}
\vspace{-.2in}
\end{figure}

\begin{IEEEproof}[Proof of Theorem~\ref{thm:core}]
Consider a $G(\beta)$ with the set of lowest degree nodes $U_l$, all having expected degrees in the interval $(w_{\min},lw_{\min})$. According to Lemma~\ref{lem:giant-component}, to ensure that the induced subgraph $G_{U_l}$ does not have a giant component, we should have
$
\sum_{v\in U_l}{w_v'}^2/\sum_{v\in U_l}{w_v'} < 1
,$
where $w_v' = w_v\frac{\sum_{v\in U_l}w_v}{N\bar{w}}$ as in Lemma~\ref{lem:subgraph}. To find $w'$, we should first obtain $\sum_{v\in U_l}w_v$. According to~\cite{chung-book}, we have 
\begin{equation}
\label{eq:volU_l}
\begin{array}{lcl}
\sum_{v\in U_l} w_v 		&\approx& N \bar{w} (1-l^{2-\beta}),\\
\sum_{v\in U_l} {w_v}^2 	&\approx& N \bar{w}^2 (1- \frac{1}{\beta-1})^2 \frac{\beta-1}{3-\beta}l^{3-\beta}.
\end{array}
\end{equation}
From~\eqref{eq:volU_l}, we conclude that $w_v'= (1-l^{2-\beta})w_v,$ for all $v\in U_l$. Thus we have,
\begin{equation}
\label{eq:volU_l-2}
\sum_{v\in U_l} w'_v \approx N \bar{w} (1-l^{2-\beta})^2
.\end{equation}
Similarly,
\begin{equation}
\label{eq:vol2U_l-2}
\sum_{v\in U_l} {w'_v}^2 \approx N \bar{w}^2 \gamma l^{3-\beta}(1-l^{2-\beta})^2
.\end{equation}
Combining~\eqref{eq:volU_l-2} and~\eqref{eq:vol2U_l-2}, we obtain the relation in~\eqref{eq:l-beta} between $\beta$ and $l$. Having $l$, we can easily obtain the size of $U_l$ by finding the number of vertices with expected degree less than $lw_{\min}$ which is readily available from~\eqref{eq:weight}. 
\end{IEEEproof}
Theorem~\ref{thm:core} provides the required information to find the size of the core and hence the number of memory units. As finding the closed-form solution for the size of the core is not straightforward, we use numerical analysis to characterize the number of required memory units using the results developed above. 
Fig.~\ref{fig:U_l} depicts the scaling of the core size versus $N$ for various $\beta$'s. As we see, the core size is a tiny fraction of the total number of nodes in the network and this fraction decreases as $N$ grows. This is a promising result as it suggests that by deploying very few memory units, we can reduce the total amount of traffic in a huge network.

\vspace{-.05in}
\section{Realization of Network Compression and Experimental Simulations}
\label{sec:routing}
\subsection{Routing in Networks Featuring Memory}
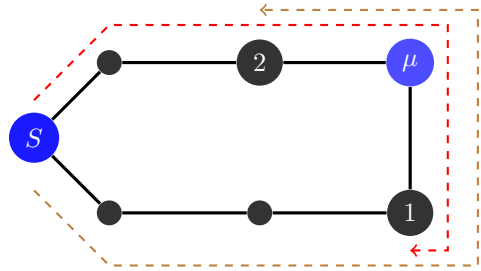
\begin{figure}
\centering
\begin{tikzpicture}
\draw (0,0) 	node[circle, fill=blue!90, text=white]	(S){$S$};
\draw (1,1) 	node[circle, fill=black!80, text=white]	(C1){{ }};
\draw (1,-1) 	node[circle, fill=black!80, text=white]	(C3){{$ $}};
\draw (3,1) 	node[circle, fill=black!80, text=white]	(C2){{$2$}};
\draw (3,-1) 	node[circle, fill=black!80, text=white]	(C4){{$ $}};
\draw (5,1) 	node[circle, fill=blue!70, text=white]	(m){{$\mu$}};
\draw (5,-1) 	node[circle, fill=black!80, text=white]	(C5){{$1$}};

\draw [very thick] (S)	--	(C1);
\draw [very thick] (S)	--	(C3);
\draw [very thick] (C1)	--	(C2);
\draw [very thick] (C2)	--	(m);
\draw [very thick] (C3)	--	(C4);
\draw [very thick] (C4)	--	(C5);
\draw [very thick] (C5)	--	(m);

\draw [->, red, thick, dashed] (0,.5)	--	(1,1.5) 	-- 	(5.5,1.5) 	-- 	(5.5,-1.5) -- (5, -1.5);
\draw [->, brown, thick, dashed] (0,-.7)	-- 	(1, -1.7)  	--	(5.9,-1.7)	--	(5.9,1.7)	--	(3,1.7);

\end{tikzpicture}
\caption{Example of routing in networks featuring memory: Memories can change the shortest paths (shown by dashed lines) dramatically. Here, $g=5$.}
\label{fig:shortestpath}
\vspace{-.2in}
\end{figure}

Thus far, we have demonstrated that a non-vanishing network-wide gain is achievable by deploying a small number of memory units in the high degree nodes in RPLG. After the memory units are deployed, the next problem that arise is the routing in networks with memory units. In networks featuring memory, the source compresses the outgoing traffic and memory nodes decompress it, and hence there exists a compression gain between source and memory. This gain of compression poses new problems for conventional routing schemes. For example, as shown in Fig.~\ref{fig:shortestpath}, let $g =5$. Then, for node 2 the cost of routing clockwise is two, but it is only $\frac{9}{5}$ ($\frac{4}{5} + 1$) by passing through memory unit $\mu$ counter-clock wise. Hence, introducing memory units in the network can dramatically change the effective shortest path. Routing is a new problem in networks featuring memory which needs to be solved differently. We consider an instance of network with a source node and fixed memory locations. We solve the routing problem in that instance of the network and after that we will be able to characterize $\mathcal{G}$.

Our approach to characterize the achievable network-wide gain is to use the results developed in Sec.~\ref{sec:RPLG-model} and compute $\mathcal{G}$ via numerical simulations. Calculation of $\mathcal{G}$ involves computing both $\mathcal{F}^0$ and $\mathcal{F}$, which in turn requires finding the shortest paths between all pairs of nodes with and without memory units. The shortest path problem in a network without memories is straightforward via Dijkstra's algorithm. But finding shortest paths in networks featuring memory requires more attention. It is important to note that in network with memory if know the shortest path from a node $u$ to $v$ and also $w$ is a node on the shortest path, this knowledge does not necessarily imply the knowledge of the shortest path from $u$ to $w$.

The Dijkstra algorithm solves the single-source shortest path for a network with positive edge costs. Our \BH~cost measure is a special case in which all the edge costs are equal to 1. However, in its original form, Dijkstra's algorithm is not applicable to networks with memory. This limitation is due to the fact that when the algorithm runs into a memory node on the path, all the previous edges' costs along the path should be divided by $g$, and hence needs to recalculate the whole path.

Here, we present a modified version of Dijkstra's algorithm that finds the effective walk from all the nodes in a network to a destination $D$, in a network with a single memory. Iterating over all nodes will provide the effective walk between every pair of nodes in the network. The extension to arbitrary number of memories is straightforward and skipped for brevity.

To handle the memory node, we define a node-marking convention by defining a set $\mathcal{M}$ which contains the marked nodes. We say that a node is marked if it is either itself a memory node, or a node through which a compressed flow is routed. The modified Dijkstra algorithm starts with finding a node $\nu$ closest to node $D$. Then, we iteratively update the effective distance of the nodes to $D$. The algorithm is summarized in Alg.~\ref{algo:Dijkstra}. The notation $cost(vD)$, used in Alg.~\ref{algo:Dijkstra}, is in fact the effective distance introduced in Sec.~\ref{sec:RPLG-model}. At the beginning, $cost(vD)=\infty$ for nodes $v$ not directly connected to $D$, and then it is calculated for $v$ in every iteration. After finding the effective distance between every pair of vertices via the modified Dijkstra algorithm, we can calculate $\mathcal{F}$ and then $\mathcal{G}$.

\begin{figure}
\begin{center}
  \includegraphics[width=.85\columnwidth]{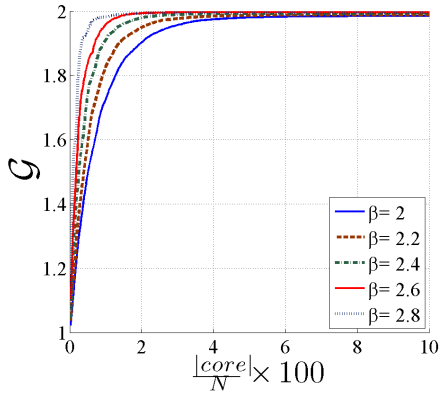}
  \caption{Illustration of the network-wide gain when simple Dijkstra routing is used. Note that $\mathcal{G}$ is capped.}
  \label{fig:G-suboptimal}
\end{center}
\end{figure}

If we do not use the modified Dijiksta algorithm in the networks with memory (i.e., we do not optimize the routing algorithm to utilize the memories), as Lemma~\ref{lem:memory-on-path} suggests, the network-wide gain would be bounded by $\frac{2g}{g+1}$. Therefore, even for very large values of $g$, the network-wide gain would remain less than two (as shown in Fig.~\ref{fig:G-suboptimal}), which is not desirable.
\begin{algorithm}
\begin{algorithmic}
\STATE $\mathcal{M}=\mathbf{\mu}$
\WHILE {$V \neq \phi$}
	\STATE $\nu$ = the closest neighbor of $D$. 
	\STATE $pathlen(\nu) = cost(\nu, D)$
\FOR {$\forall v \in V\setminus \{\nu,D\}$}
        \IF {$\nu \not\in \mathcal{M}$}
        	\STATE {$cost(vD) = \min\{cost(vD),cost(v\nu) + cost(\nu D)\}$}
        \ELSE
        	\STATE $cost(vD) = \min\{cost(vD),\frac{cost(v\nu)}{g} + cost(\nu D)\}$
        	\STATE $\mathcal{M}\gets\mathcal{M} \cup v$
        \ENDIF
\ENDFOR
\STATE $V\gets V \setminus \nu$
\ENDWHILE
\end{algorithmic}
\caption{Modified Dijkstra's Algorithm}
\label{algo:Dijkstra}
\end{algorithm}
\vspace{-.1in}

\vspace{-.05in}
\subsection{Simulation Results}
To validate our theoretical results, we have conducted different sets of experiments to characterize the network-wide gain of memory in RPLG. For experiments, we used DIGG RPLG generator~\cite{Brady2006}, with which we generated random power-law graph instances with number of vertices between 1000 and 5000, and $2<\beta<3$. The result are averaged over 5 instances of generated RPLG. In our simulations, we report results for various core sizes (number of memory units).

\begin{figure}
\begin{center}
  \includegraphics[width=.85\columnwidth]{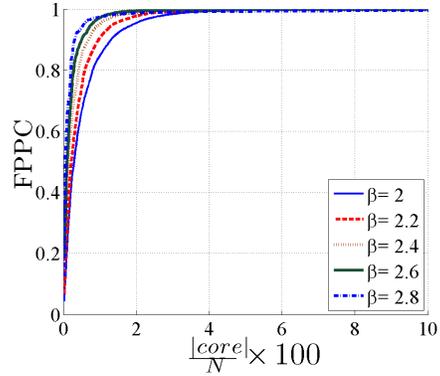}
  \caption{The fraction of the paths passing through the core (FPPC) vs. the core size, for RPLG of size $N=5000$.}
  \label{fig:path-through-core}
\end{center}
\end{figure}

We first verify our assumption that a tiny fraction of the highest degree nodes observes most of the traffic in the network. Fig.~\ref{fig:path-through-core} shows the fraction of the paths that pass through the core (FPPC) for different core sizes and $\beta$'s. As we expected, more that $90\%$ of the shortest paths in the graph involves less than $2\%$ of the highest degree nodes to route the flow. Although our theoretical result in Theorem~\ref{thm:core} is asymptotic in $N$, Fig.~\ref{fig:path-through-core} suggests that our result holds for moderate values of $N$ as well. Therefore, we can place the memory units at the core and results can be extrapolated for large graphs with large number of nodes.

\begin{figure*}
\begin{center}
  \subfigure[$\frac{|core|}{N}=2.5\%$]{
  \includegraphics[width=.54\columnwidth]{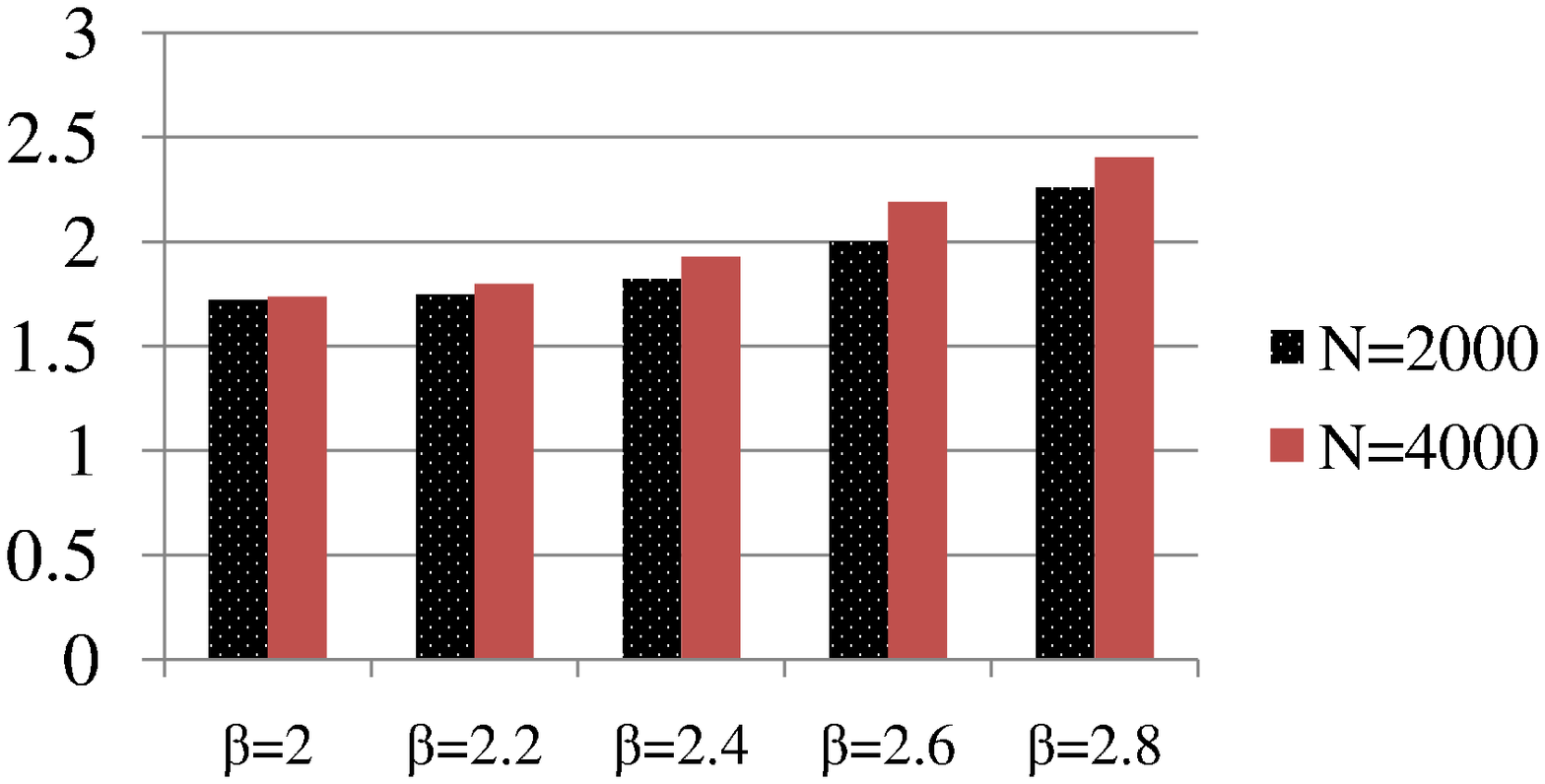}
  \label{fig:G-mem-2-5}
  }
  \subfigure[$\frac{|core|}{N}=5\%$]{
  \includegraphics[width=.54\columnwidth]{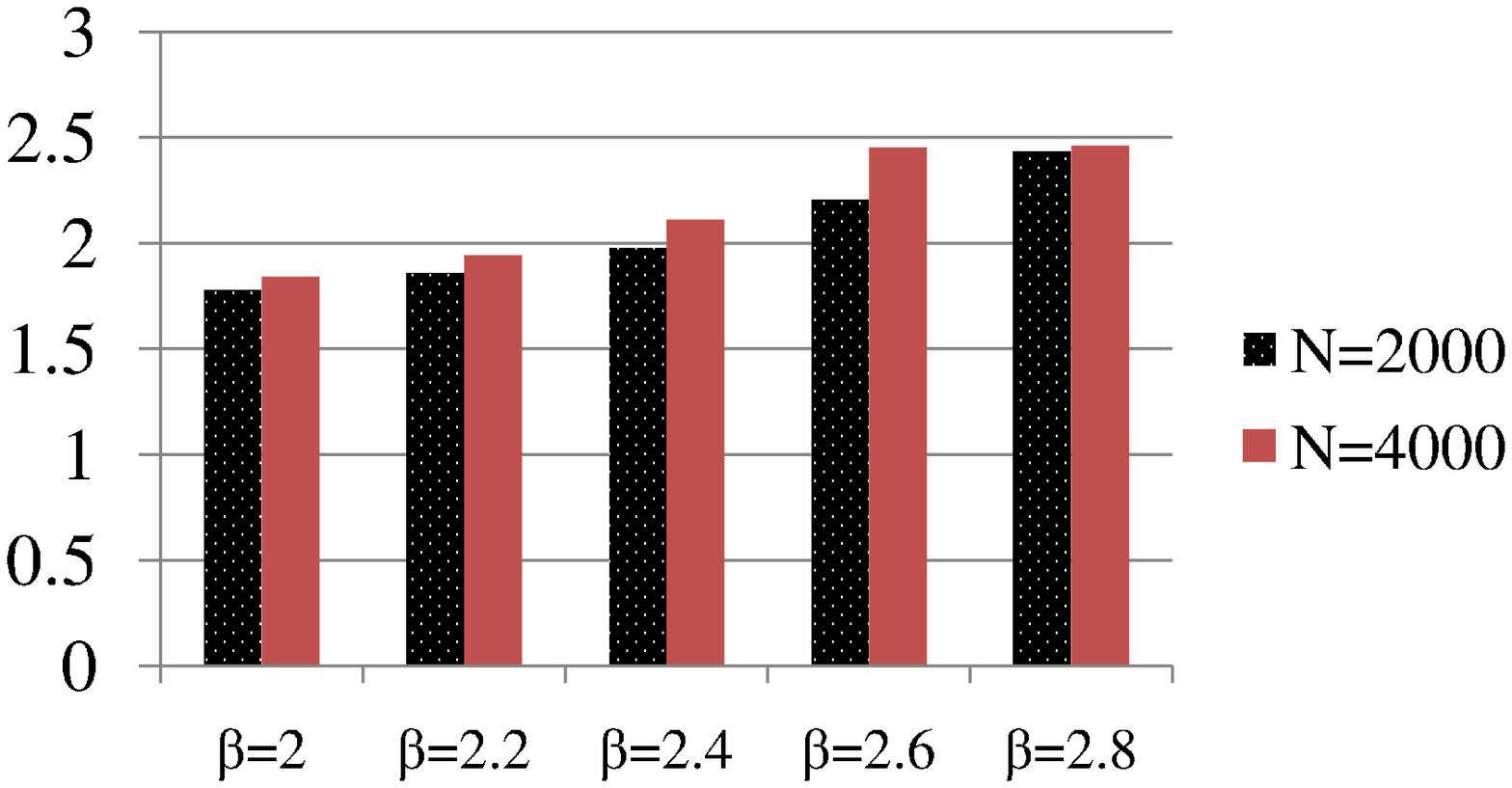}
  \label{fig:G-mem-5}
  }
  \subfigure[$\frac{|core|}{N}=10\%$]{
  \includegraphics[width=.54\columnwidth]{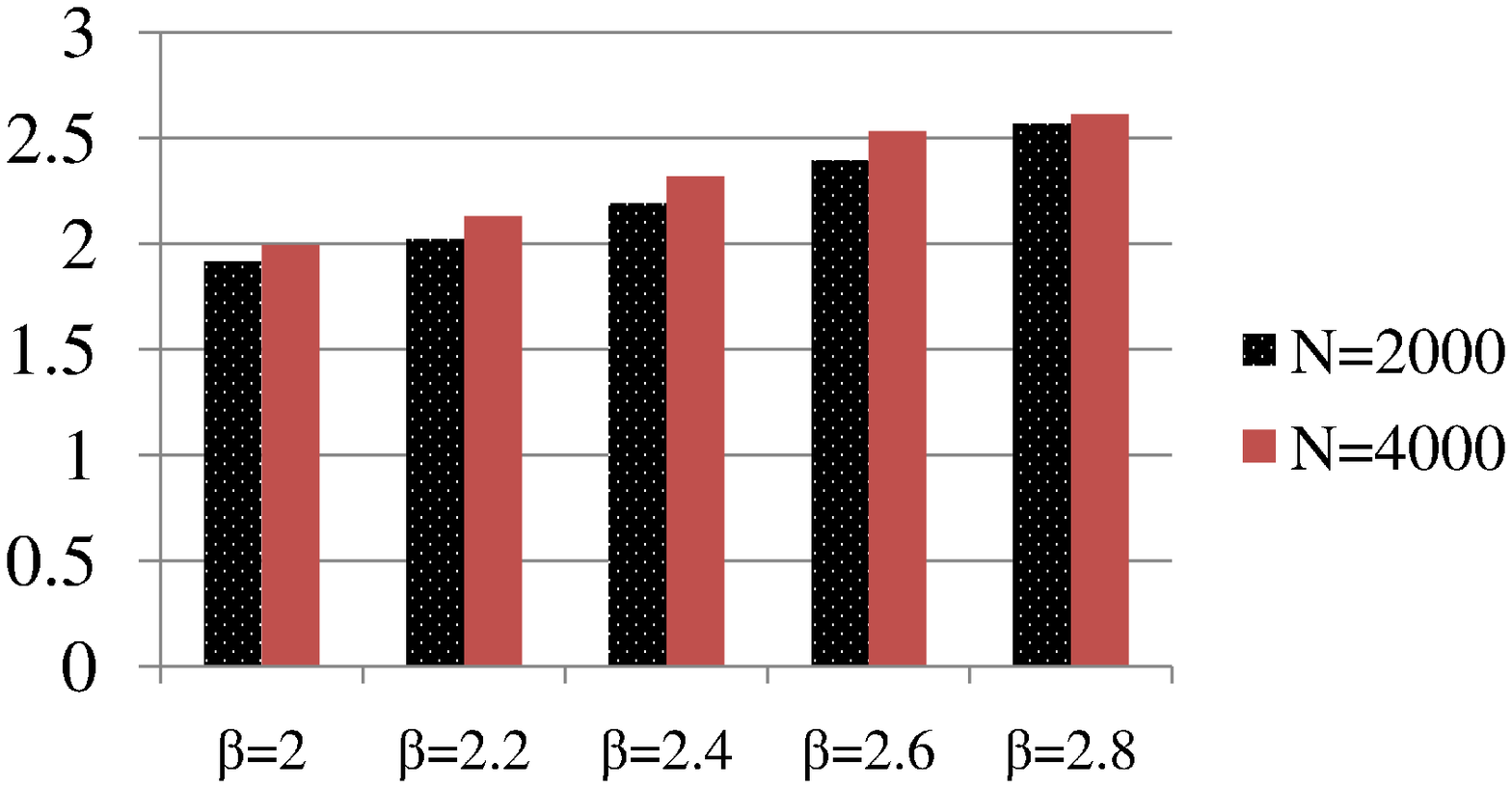}
  \label{fig:G-mem-10}
  }
  \subfigure{
  \includegraphics[width=.15\columnwidth]{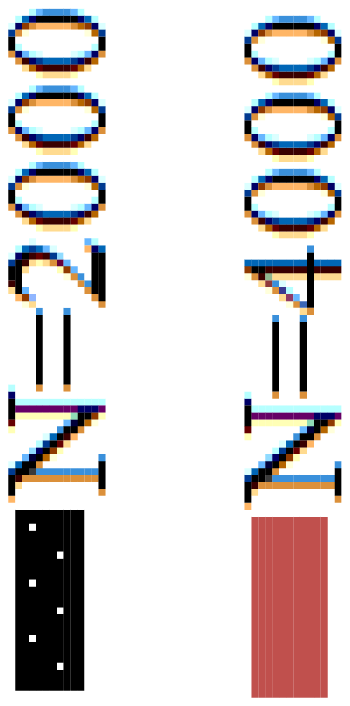}
  }
\end{center}
\vspace{-.1in}
\caption{Network-wide gain of memory assisted compression $\mathcal{G}$ for different core sizes and power-law parameter $\beta$, for $g=3$. }
\label{fig:network-wide-gains}
\vspace{-.1in}
\end{figure*}

To validate the network-wide gain of memory, we have considered two RPLGs with sizes $N=2000$ and $N=4000$. Assume that each memory node has observed a sequence of length $m=4$MB of previous communications in the network. The packets transmitted in the network are of size 1kB. This assumption is in accordance with the Maximum Transmission Unit (MTU) of 1500-bytes allowed by Ethernet at the network layer. From our results in Sec.~\ref{sec:memory-assisted}, we observe that a memory-assisted source compression gain of $g\approx 2.5$ is achievable for real traffic traces. Hence, we use $g=3$ in our simulations.

Fig.~\ref{fig:network-wide-gains} describes our results for the achievable network-wide gain of memory-assisted compression. We measured the total flow without memory. We also obtain the optimal paths when we have memory units are deployed. We consider three cases in which the fraction pf nodes equipped with memory increases from $2.5\%$ of the nodes to $10\%$. All data has been averaged over the 5 graphs in each set.

The trendlines suggest that $\mathcal{G}$ increases as $\beta$ increases which is expected since the FPPC increases with $\beta$. In other words, more traffic between the nodes in periphery has to travel through the dense subgraph (core) as $\beta$ increases. Further, by increasing the number of memory units, the network-wide gain increases and approaches to the upper bound $g$. It is important to note that enabling only $2.5\%$ of the nodes in the network with memory-assisted compression capability, we can reduce the total traffic in the network by a factor of $ 2$
on top of flow compression without using memory, i.e., end-to-end compression. We emphasize that this memory-assisted compression (UcompM) feature does not have extra computation overhead for the source node (in comparison with the end-to-end compression technique in Ucomp). Further, this feature only requires extra computation at the memory units when compared to Ucomp. However, this extra computation incurs a linear complexity with the length of the data traffic. Hence, overall with some additional linear computational complexity on top of what could have been achieved using a mere end-to-end compression, the memory-assisted compression can reduce the traffic by a factor of $2$.

\vspace{-.05in}
\section{Conclusion}
\label{sec:conclusion}
In this paper, we employed the concept of memory-assisted compression of information sources and introduced its implication in reducing the amount of traffic in real networks.
The basic idea is to allow some intermediate nodes in the network to be capable of memorization and compression. The memory units observe the traffic of the network and form a model for the information source. Then, using the source model a better universal compression of the flow is achieved in the network.
We presented two algorithms for the memory-assisted compression of data content that do not require any prior knowledge of the statistical distribution of the content. We demonstrated the effectiveness of both algorithms via simulation on real data traces and characterized the memorization gain on a single link. Then, we investigated, from information theory point of view, the network flow compression by utilizing memory units in the network. We considered Internet-like power-law graphs to develop theoretical results on the number of memory units needed to obtain network-wide gain and, as an intermediate step,  solved the routing problem for networks with memory units. We showed that non-vanishing network-wide gain of memorization is obtained even when the number of memory units is a tiny fraction of the total number of nodes in the network. Finally, we validated our predictions of the memorization gain by simulation on real traffic traces. We observed that by enabling compression on less than $2\%$ of the nodes, we can expect network-wide gains of very close to its upper-bound of $g$. That is more than two-fold gain over conventional end-to-end compression.


\vspace{-.05in}
\bibliographystyle{IEEEtran}
\bibliography{net-comp}
\end{document}